\documentclass[12pt,preprint,flushrt]{aastex}
\begin{document}
\newcommand{\etal}{{\it et al.}}
\include{idl}

\title{Detection of Nitrogen and Neon in the X-ray Spectrum of GP Com
with {\it XMM/Newton}} \author{Tod E. Strohmayer} \affil{Laboratory
for High Energy Astrophysics, NASA's Goddard Space Flight Center,
Greenbelt, MD 20771; stroh@clarence.gsfc.nasa.gov}

\begin{abstract}

We report on X-ray spectroscopic observations with {\it XMM/Newton} of
the ultracompact, double white dwarf binary, GP Com. With the
Reflection Grating Spectrometers (RGS) we detect the L$\alpha$ and
L$\beta$ lines of hydrogen-like nitrogen (N VII) and neon (Ne X), as
well as the helium-like triplets (N VI and Ne IX) of these same
elements. All the emission lines are unresolved. These are the first
detections of X-ray emission lines from a double-degenerate, AM CVn
system. We detect the resonance (r) and intercombination (i) lines of
the N VI triplet, but not the forbidden (f) line. The implied line
ratios for N VI, $R = f/i < 0.3$, and $G = (f + i)/r \approx 1$,
combined with the strong resonance line are consistent with formation
in a dense, collision-dominated plasma. Both the RGS and EPIC/MOS
spectra are well fit by emission from an optically thin thermal plasma
with an emission measure (EM) $\propto ( kT / 6.5 \; {\rm keV})^{0.8}$
(model {\it cevmkl} in XSPEC). Helium, nitrogen, oxygen and neon are
required to adequately model the spectrum, however, the inclusion of
sulphur and iron further improves the fit, suggesting these elements
may also be present at low abundance. We confirm in the X-rays the
underabundance of both carbon and oxygen relative to nitrogen, first
deduced from optical spectroscopy by Marsh et al. The average X-ray
luminosity of $\approx 3 \times 10^{30}$ ergs s$^{-1}$ implies a mass
accretion rate $\dot m \approx 9 \times 10^{-13} \; M_{\odot}\; {\rm
yr}^{-1}$. The implied temperature and density of the emitting plasma,
combined with the presence of narrow emission lines and the low $\dot
m$ value, are consistent with production of the X-ray emission in an
optically thin boundary layer just above the surface of the white
dwarf.

\end{abstract}


\section{Introduction}

The AM CVn stars are among the most compact binary systems
known. Their optical spectra are dominated by broad helium emission
lines originating in an accretion disk formed by mass transfer from a
degenerate helium dwarf onto the primary white dwarf. Their orbital
periods range from about 10 minutes to 1 hour.  They are natural
laboratories for the study of poorly understood binary evolution
processes, such as common envelope evolution.  Moreover, the absence
of hydrogen allows for the study of accretion disks dominated by
helium. They may also be a significant channel for the production of
Type Ia supernovae and neutron stars via accretion induced collapse
(for a review see Warner 1995).

GP Com is one of the better studied, nearby AM CVn systems. It has a
46.5 minute orbital period (Nather Robinson \& Stover 1981), resides
at a distance of $\approx 70$ pc (Thorstensen 2003), and has been
studied extensively in the optical and UV. Recent studies by Marsh
(1999) and Morales-Rueda et al.(2003) have used optical and UV+optical
spectroscopy to probe the dynamics of the system in great detail. In
addition to helium, nitrogen emission lines have been seen in the UV
(Lambert \& Slovak 1981; Marsh et al. 1995), and optical (Marsh, Horne
\& Rosen 1991). These observations identified a nitrogen
overabundance, relative to carbon and oxygen, as well as an
underabundance of heavy elements (for example, Ca, Si and Fe). The low
metallicity may be consistent with its suggested halo origin (see
Giclas, Burnham \& Thomas 1961). The evolution scenario favored by
Marsh et al. (1991) posits an initially low metallicity for the
progenitor stars. The CNO elements are subsequently produced within
the primary star and then mixed throughout the secondary during a
common envelope phase. The nitrogen abundance is then increased by
CNO-cycle hydrogen burning within the secondary (ie. the mass donor).
Finally, a number of ultracompact systems with neutron star primaries
also show apparent neon enrichment (see Juett \& Chakrabarty 2003;
Schulz et al. 2001).

Accretion should make such systems soft X-ray sources, and indeed,
several have been detected in the X-ray band, including the prototype
AM CVn, CR Boo, and GP Com (van Teeseling \& Verbunt 1994; Ulla 1995;
and Eracleous, Halpern \& Patterson 1991). The X-ray flux from
accreting, non-magnetic white dwarfs is thought to originate in a
boundary layer that shocks and decelerates the Keplerian flow,
allowing it to eventually settle on the white dwarf. Early work
suggested that the boundary layer should be optically thick--radiating
in the soft X-ray band and extreme UV--at mass accretion rates above
$\dot m \approx 1.6 \times 10^{-10} M_{\odot} \; {\rm yr}^{-1}$
(Pringle \& Savonije 1979). At lower rates the boundary layer should
become more radially extended and optically thin, producing a harder
X-ray spectrum. Aspects of this basic scenario have been confirmed by
more recent calculations (see, for example, Patterson \& Raymond 1985;
Narayan \& Popham 1993; Popham \& Narayan 1995).

To date, X-ray observations of most AM CVn systems have been made at
relatively low signal to noise levels and with modest spectral
resolution. For example, ROSAT observations of GP Com found the source
at a flux level $\approx 1.2 \times 10^{-11}$ ergs cm$^{-2}$ s$^{-1}$
in the 0.2 - 2 keV band, and confirmed that it is variable. These
observations also found that the 0.2 - 2 keV spectrum cannot be
described by a single temperature component (van Teeseling \& Verbunt
1994), but detailed modelling was problematic due to the relatively
poor statistics and restricted bandpass.

In this Letter we summarize results of recent, high signal to noise
X-ray spectral measurements of GP Com with {\it XMM/Newton}.  These
data represent perhaps the best X-ray spectral measurements of an AM
CVn system to date, and also show the first X-ray emission lines from
such a system. 

\section{XMM Observations} 

XMM/Newton observed GP Com for 56.5 ksec beginning on January 3,
2001. The EPIC observations were done in imaging mode. We used version
5.4.1 of the XMM software (SAS) to analyze the data. GP Com was easily
detected at the few counts s$^{-1}$ level in the EPIC instruments. A
lightcurve combining data from the PN and MOS detectors is shown in
Figure 1, and confirms substantial X-ray variability, including
several 2 to 4 minute flares during which the X-ray flux increased by
factors of 2 - 3. Flaring has often been seen in the UV and optical,
particularly in the N V ($1240 \AA$) resonance line (Marsh et
al. 1995). We will present a detailed timing study of GP Com in a
subsequent paper.  We extracted spectra and produced response files
for each of the instruments using the relevant SAS tools. 
%For the purposes of this work we present results on only the total,
%phase-averaged spectrum.

\subsection{Spectral Analysis: RGS and EPIC/MOS}

The spectrum extracted from both RGS instruments is shown in Figure 2,
along with identifications for all the strongest line features.  The
strongest detected feature is the Ly$\alpha$ line of hydrogenic
nitrogen (N VII) at 24.78 $\AA$, but other lines of nitrogen are also
present, including the Ly$\beta$ line as well as the helium-like
triplet of N VI at $\approx 29 \AA$. The only other signifcant line
features are due to neon. The hydrogenic Ly$\alpha$ and Ly$\beta$
lines as well as the helium-like triplet are detected. Particularly
striking is the absence of any strong carbon or oxygen lines in the
spectrum.

We fit the observed line features using Lorentzian profiles with the
widths fixed at that of the RGS line spread function (LSF) for that
wavelength (see den Herder et al. 2001; Pollock et al. 2003). All the
hydrogenic lines are well fit with such a profile and are
unresolved. Table 1 summarizes the salient properties of all the
detected lines.

As is well known, the helium-like triplets can be used as probes of
the plasma where the lines are produced (see Gabriel \& Jordan 1969;
Porquet \& Dubau 2000). The line flux ratios, $R = f/i$, and $G = (i +
f)/r$, of the resonance (r), intercombination (i) and forbidden (f)
transitions are sensitive to the electron density and temperature of
the plasma, respectively. For each triplet we fit a Lorentzian profile
at the rest wavelength of each transition. To determine the line
ratios, we fixed the wavelength and width of each line, but allowed
the line fluxes to vary.  The signal to noise ratio in both triplets
is not particularly high, but the line ratios can at least be roughly
estimated. The nitrogen triplet is better resolved than that of
neon. The results of our fits are shown in Figure 3 (see also Table
1). For the N VI triplet we detect the resonance and intercombination
lines, but interestingly, not the forbidden line. The line ratios for
N VI, $R = f/i < 0.3$, and $G = (f + i)/r \approx 1$, and the strong
resonance line are consistent with formation in a dense,
collision-dominated plasma (Porquet \& Dubau 2000). For nitrogen, a
strong UV radiation field can also decrease the strength of the
forbidden line, mimicking a density effect, however, the line ratios
from the neon triplet--which are much less sensitive to radiation
field effects--are also consistent with a plasma at or near the
critical density $n_c = 5.3 \times 10^{9}$ cm$^{-3}$ for nitrogen.

We fit the EPIC/MOS and RGS spectra with several collisional
ionization equilibrium (CIE) plasma emission models. We found that
multi-temperature models fit the data significantly better than single
temperature models. For example, the multi-temperature and variable
abundance model {\it cevmkl} in XSPEC (version 11.3) provides an
excellent fit to the data from both the RGS and EPIC instruments (see
Mewe, Gronenschild \& van den Oord 1985; Mewe, Lemen \& van den Oord
1986; Liedahl, Osterheld \& Goldstein 1995). This model uses a power
law in temperature for the emission measure distribution (EM), vis.,
${\rm EM} \propto ( kT/ kT_{max} )^{\gamma}$, where $kT_{max}$ and
$\gamma$ are the maximum temperature and power law index,
respectively.

There is some evidence for modest pile-up in the PN spectra, so for
the broad-band modeling we focused on the MOS data only. Figure 4
shows our best fit to the EPIC/MOS data using the {\it cevmkl} model.
The model fits the data well, with a minimum $\chi^2$ of 1,005 for
1,045 degrees of freedom. Using the MOS data only, the best-fit
emission measure distribution is characterized by $kT_{max} = 6.3 \pm
0.3$ keV, and $\gamma = 0.85 \pm 0.05$. The 0.3 - 10 keV fluxes are
$4.7$ and $5.3 \times 10^{-12}$ ergs cm$^{-2}$ s$^{-1}$ in the MOS1
and MOS2, respectively.  Taking the mean, we arrive at an average
luminosity (at 70 pc) of $2.94 \times 10^{30}$ ergs s$^{-1}$ in the
same band.

The elements strongly required by the fit are helium, nitrogen, oxygen
and neon, however, $\chi^2$ additionally improves by 13.35 and 23.9
with the inclusion of iron and sulphur, respectively, suggesting these
elements may also be present at low abundance. No other elements are
required to fit the spectrum.  Using this model we determined the
abundances relative to the solar values tabulated by Anders \&
Grevesse (1989). Expressed as mass fractions we find; $X_{He} = 0.977
\pm 0.002$, $X_{N} = 1.7 \pm 0.1 \times 10^{-2}$, $X_{O} = 2.2 \pm 0.3
\times 10^{-3}$, $X_{Ne} = 3.7 \pm 0.2 \times 10^{-3}$, $X_{S} = 2.3
\pm 0.6 \times 10^{-4}$, and $X_{Fe} = 8 \pm 2 \times 10^{-5}$. In
addition, we placed limits ($3\sigma$) on several prominent elements
which were not detected; $X_{C} < 2 \times 10^{-3}$, $X_{Mg} < 1.6
\times 10^{-4}$ and $X_{Si} < 8 \times 10^{-5}$. Our limits on some of
the heavier metals (such as Si, Mg, and Fe) suggest underabundances
compared with solar values.  This is qualitatively similar to the
abundances derived from optical spectroscopy (Marsh et al. 1991).

\section{Discussion}

The nitrogen lines in the X-ray band provide additional confirmation
of the overabundance of nitrogen relative to carbon and oxygen deduced
from the optical (Marsh et al. 1991). The high nitrogen abundance
likely reflects the operation of CNO-cycle thermonuclear processing in
the secondary (Marsh et al. 1991).  We find a nitrogen to helium
ratio, by number, of $\approx 4.8 \times 10^{-3}$, which is moderately
higher than the value derived by Marsh et al. (1991) of $2.7 \times
10^{-3}$ (see their Table 4). Our derived oxygen to helium number
ratio is, however, about a factor of 10 higher than that deduced from
the optical. Although systematic modelling effects could be present,
these results suggest that the X-ray emitting gas may have higher
oxygen abundance than the accretion disk (which produces the optical
emission lines). If the accretor were an oxygen-rich white dwarf, and
some of its matter is mixed with the accreted material in the
turbulent boundary layer, then this could perhaps increase the oxygen
abundance relative to the donor material.

The neon lines provide the first definitive detection of an element
heavier than oxygen in GP Com. Our derived neon mass fraction
($\approx 0.0037$) is a factor of 2 larger than the solar
value. Moreover, modelling of the optical spectra also suggests a neon
enrichment of about 1.3 compared to solar abundances (Marsh et al.
1991). Neon is a by-product of helium burning under a wide range of
conditions (see, for example, Clayton 1983; Iben \& Truran 1983),
however, the low carbon abundance is likely incompatible with helium
burning having ocurred in the secondary. Therefore, this avenue for
the neon production appears to be a dead-end.  Again, mixing of neon
from the accretor (which likely has undergone helium burning) into the
donor material in the boundary layer could perhaps increase the neon
abundance.  It appears that gravitational settling and fractionation
of neon (in the form of $^{22}Ne$) are also not relevant, since
nitrogen is depleted in the production of this isotope, and we clearly
measure a large nitrogen abundance (see, for example, Bildsten \& Hall
2001).

Neon enhancements have been suggested for several ultracompact neutron
star binaries; 4U 1543-624 and 2S 0918-549 (Juett \& Chakrabarty
2003), and 4U 1627-67 (Schulz et al. 2001). For 4U 1543-624, Juett \&
Chakrabarty (2003) deduced Ne/O $\approx 1.5$, by number, from studies
of absorption edges in {\it Chandra} high resolution spectra, however,
we note that these measurements can be influenced by the unknown
ionization structure in the circumbinary material.  We find a similar
value of Ne/O = 1.4 (by number) for GP Com. Yungelson, Nelemans \& van
den Heuvel (2002) suggest fractionation of neon in the white dwarf
donor as a possible explanation for the neon enhancements in the
neutron star systems. This scenario again requires a helium burning
donor and thus seems not to work in the case of GP Com.

The X-ray measurements provide a means to estimate the mass accretion
rate, $\dot m$, onto the white dwarf primary in GP Com. Assuming
matter falls from the inner Lagrange point onto the primary, and that
half the gravitational potential energy is dissipated in the accretion
disk, the X-ray luminosity, $L_x$ can be approximated as (see, for
example, Nelemans, Yungelson \& Portegies Zwart 2004),
\begin{equation}
L_x = \frac{1}{2}\frac{G M_1 \dot m}{R} \left (1 - \frac{R}{R_{L_1}}
\right ) \; ,
\end{equation}
where $M_1$, $R$, $R_{L_1}$, and $\dot m$ are the primary mass,
primary radius, the distance from the inner Lagrange point to the
center of the accretor, and the mass accretion rate,
respectively. Solving for $\dot m$, and inserting the measured X-ray
luminosity for $L_x$ gives,
\begin{equation}
\dot m = 3.48 \times 10^{-13} \frac{R_9}{(M_1/M_{\odot})} \left ( 1 -
\frac{R}{R_{L_1}} \right )^{-1} \; \; M_{\odot} \; {\rm yr}^{-1} \; ,
\end{equation}
where $R_9$ is the radius of the primary in units of $10^9$ cm, and
$R/R_{L_1} \approx 0.2$ for a system like GP Com. For a primary mass,
$M_1 = 0.5 M_{\odot}$ (Morales-Rueda et al. 2003) and $R_9 = 1$
(Zapolsky \& Salpeter (1969) we obtain $\dot m \approx 8.7 \times
10^{-13} \; M_{\odot} \; {\rm yr}^{-1}$. If the accretor is a massive
O/Ne/Mg white dwarf (suggested by the possibility of oxygen and neon
mixing in the boundary layer), then the derived accretion rate drops
by about a factor of 6.

This rate is substantially lower than the critical rate above which
one expects an optically thick boundary layer (Pringle \& Savonije
1979; Narayan \& Popham 1993). In this regime the plasma should have a
maximum temperature of $\approx 10^8$ K, which is similar to the peak
temperature of $6.5$ keV ($7.5 \times 10^7$ K) deduced from our
spectral modelling.  The presence of {\it narrow} emission lines also
supports spectral formation in a boundary layer. From the width of the
N VII Ly$\alpha$ line, we obtain an upper limit on the velocity
dispersion of the emitting gas of $\approx 250$ km s$^{-1}$
($3\sigma$) (This includes bulk and thermal motions).  This indicates
an origin close to the white dwarf and not, for example, further out
in the accretion disk, which would have much higher Keplerian
velocities.  If the white dwarf accretor was synchronized with the
orbit, then its rotational velocity would be $\approx 23$ km
s$^{-1}$. To have a rotational speed equal to our 250 km s$^{-1}$
limit, a $10^{4}$ km white dwarf would require a spin period of
$\approx 250$ seconds. Indeed, an upper limit to the Doppler
(ie. thermal motion) width for the nitrogen line indicates a value
close to t250 km s$^{-1}$, which supports a slowly rotating, or
synchronized white dwarf. The narrowness of the X-ray lines also
appears consistent with the narrow ``central spike'' component of the
helium lines (Marsh 1999; Morales-Rueda et al. 2003), which almost
certainly originate on the accreting white dwarf.

The XMM/Newton measurements provide a detailed look at the physics of
the boundary layer in GP Com. Detailed comparisons with theoretical
boundary layer models could provide interesting constraints on their
temperature, density and rotational profiles. Deeper X-ray spectra
would provide better temperature and density constraints from the
helium-like triplets, and likely more lines of less abundant species
could be detected.

\acknowledgements

The author thanks Richard Mushotzky, Martin Still, Craig Markwardt,
and the referee, Tom Marsh, for many useful comments and suggestions.

\centerline{\bf References}

\noindent{} Anders E. \& Grevesse N. 1989, Geochimica et Cosmochimica
Acta 53, 197.

\noindent{} Bildsten, L. \& Hall, D. M. 2001, ApJ, 549, L219.

\noindent{} Clayton, D. D. 1983, ``Principles of Stellar Evolution and
Nucleosynthesis,'' (Univ. of Chicago Press: Chicago).

\noindent{} Eracleous, M., Halpern, J. \& Patterson, J. 1991, ApJ, 382, 290.

\noindent{} den Herder, J. W. et al. 2001, A\&A, 365, L7.

\noindent{} Gabriel, A. H. \& Jordan, C. 1969, MNRAS, 145, 241.

\noindent{} Giclas, H. L., Burnham, R., Jr., \& Thomas, N. G. 1961, Lowell
Obs. Bull., 112, 61.

\noindent{} Iben, I. \& Truran, J. W. 1978, ApJ, 220, 980.

\noindent{} Juett, A. M. \& Chakrabarty, D. 2003, ApJ, 599, 498.

\noindent{} Lambert, D. L. \& Slovak, M, H. 1981, PASP, 93, 477.

\noindent{} Liedahl, D.A., Osterheld, A.L., and Goldstein, W.H. 1995,
ApJL, 438, 115

\noindent{} Marsh, T. R., Wood, J. H., Horne, K. \& Lambert, D. 1995, MNRAS, 
274, 452.

\noindent{} Marsh, T. R., Horne, K. \& Rosen, S. 1991, ApJ, 366, 535.

\noindent{} Marsh, T. R. 1999, MNRAS, 304, 443.

\noindent{} Mewe, R., Gronenschild, E.H.B.M., \& van den Oord, G.H.J. 1985, 
A\&AS, 62, 197.

\noindent{} Mewe, R., Lemen, J.R., \& van den Oord, G.H.J. 1986,
A\&AS, 65, 511.

\noindent{} Morales-Rueda, L. et al. 2003, A\&A, 405 249,

\noindent{} Narayan, R. \& Popham, R, G. 1993, Nature, 362, 820.

\noindent{} Nather, R. E., Robinson, E. L. \& Stover, R. J. 1981, ApJ,
244, 269.

\noindent{} Nelemans, G., Yungelson, L.~R., \& Portegies Zwart, S.~F.\
2004, MNRAS, 349, 181.

\noindent{} Patterson, J. \& Raymond, J. C. 1985, ApJ, 292, 535.

\noindent{} Pollock, A. M. T. et al. 2003, XMM Calibration Document,
XMM-SOC-CAL-TN-0030 Issue 2, ``Status of the RGS Calibration,''

\noindent{} Popham, R. G. \& Narayan, R. 1995, ApJ, 442, 337.

\noindent{} Porquet, D. \& Dubau, J. 2000, A\&AS, 143, 495.

\noindent{} Pringle, J. E. \& Savonije, G. J. 1979, MNRAS, 187, 777.

\noindent{} Schulz, N. S. et al. 2001, ApJ, 563, 941.

\noindent{} Thorstensen, J. R. 2003, AJ, 126, 3017.

\noindent{} Ulla, A. 1995, A\&A, 301, 469.

\noindent{} van Teeseling, A. \& Verbunt, F. 1994, A\&A, 292, 519.

\noindent{} Warner, B. 1995, {\it Cataclysmic Variable Stars}, Cambridge
Univ. Press, Cambridge UK.

\noindent{} Yungelson, L. R., Nelemans, G. \& van den Heuvel,
E. P. J. 2002, A\&A, 388, 546.

\noindent{} Zapolsky, H. S. \& Salpeter, E. E. 1969, ApJ, 158, 809.

\pagebreak

\begin{figure}
\begin{center}
\includegraphics[width=6in, height=6in]{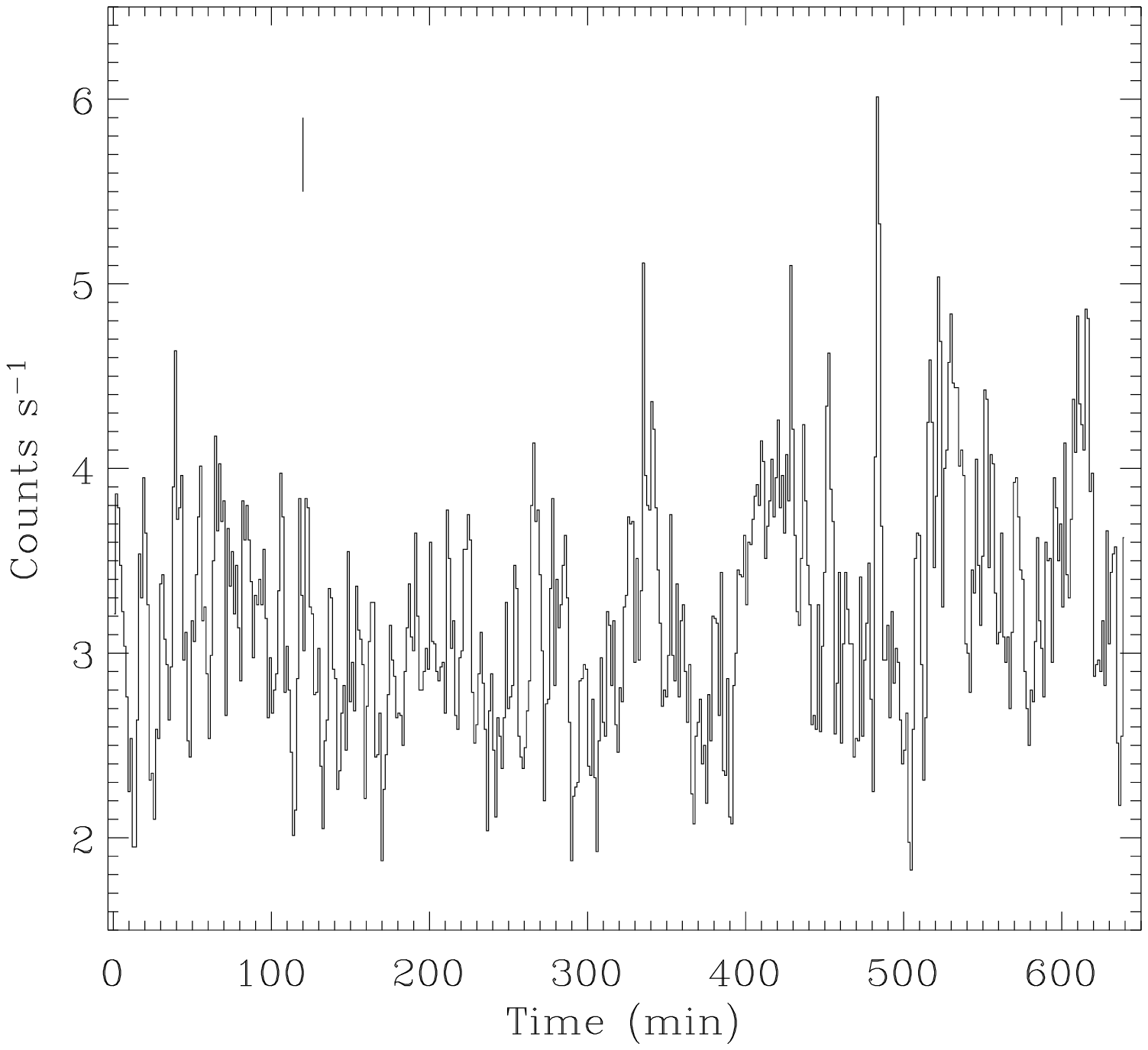}
\end{center}
Figure 1: A lightcurve of a portion of the {\it XMM/Newton} EPIC
(PN+MOS) observation of GP Com. The binsize is 80 s. Significant
variability is evident on a range of timescales. A characteristic
error bar is also shown (upper left).
\end{figure}
\clearpage

\begin{figure}
\begin{center}
 \includegraphics[width=6in, height=6in]{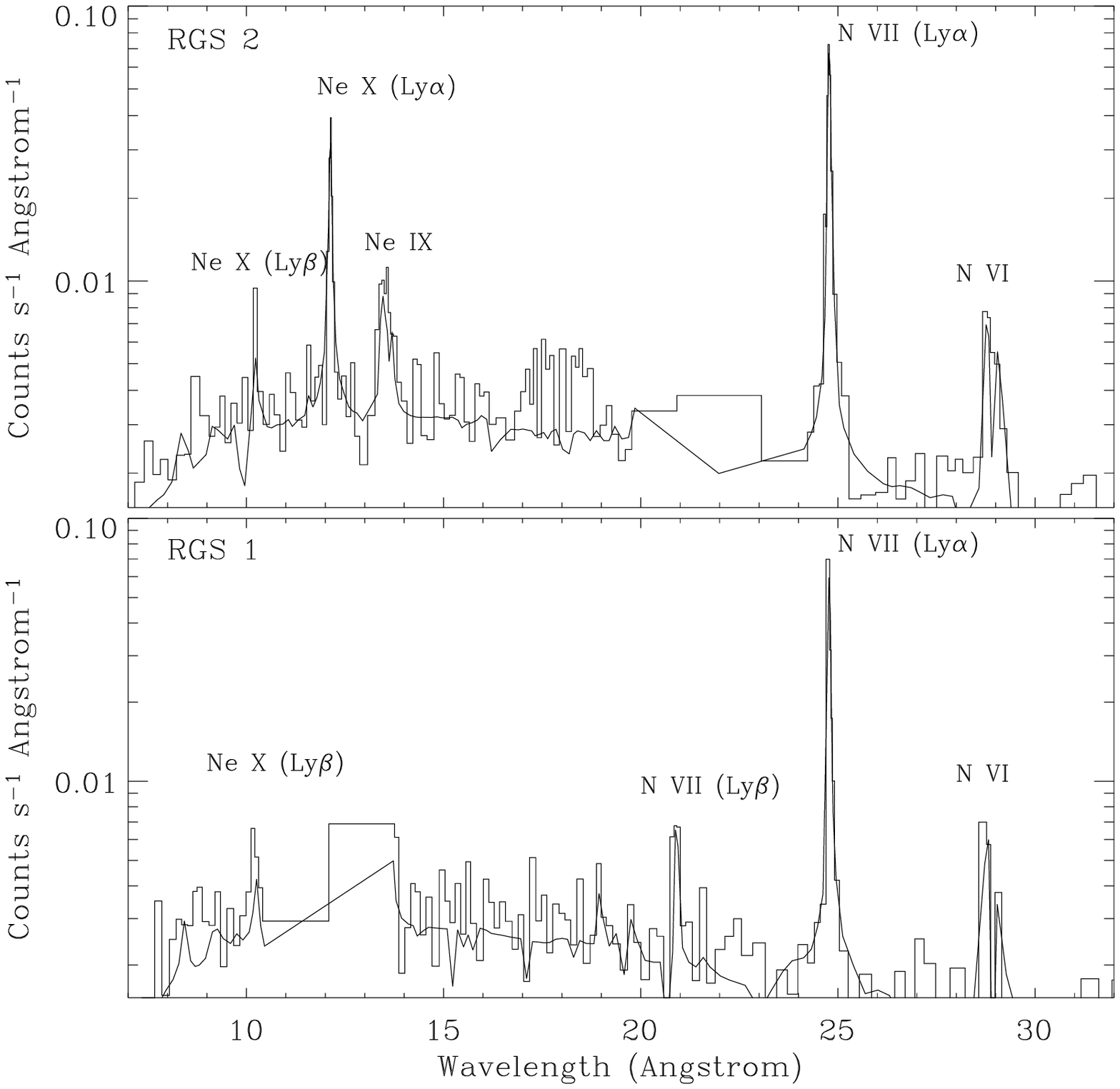}
\end{center}
Figure 2: High resolution spectra of GP Com obtained with the RGS1
(bottom) and RGS2 (top) spectrometers in the $7 - 32 \; \AA$ band
(solid histogram). Identifications are given for the prominent
lines. The best fitting, variable emission measure plasma model
(cevmkl in XSPEC) is also shown (solid curve).
\end{figure}
\clearpage

\begin{figure}
\begin{center}
 \includegraphics[width=6in, height=5in]{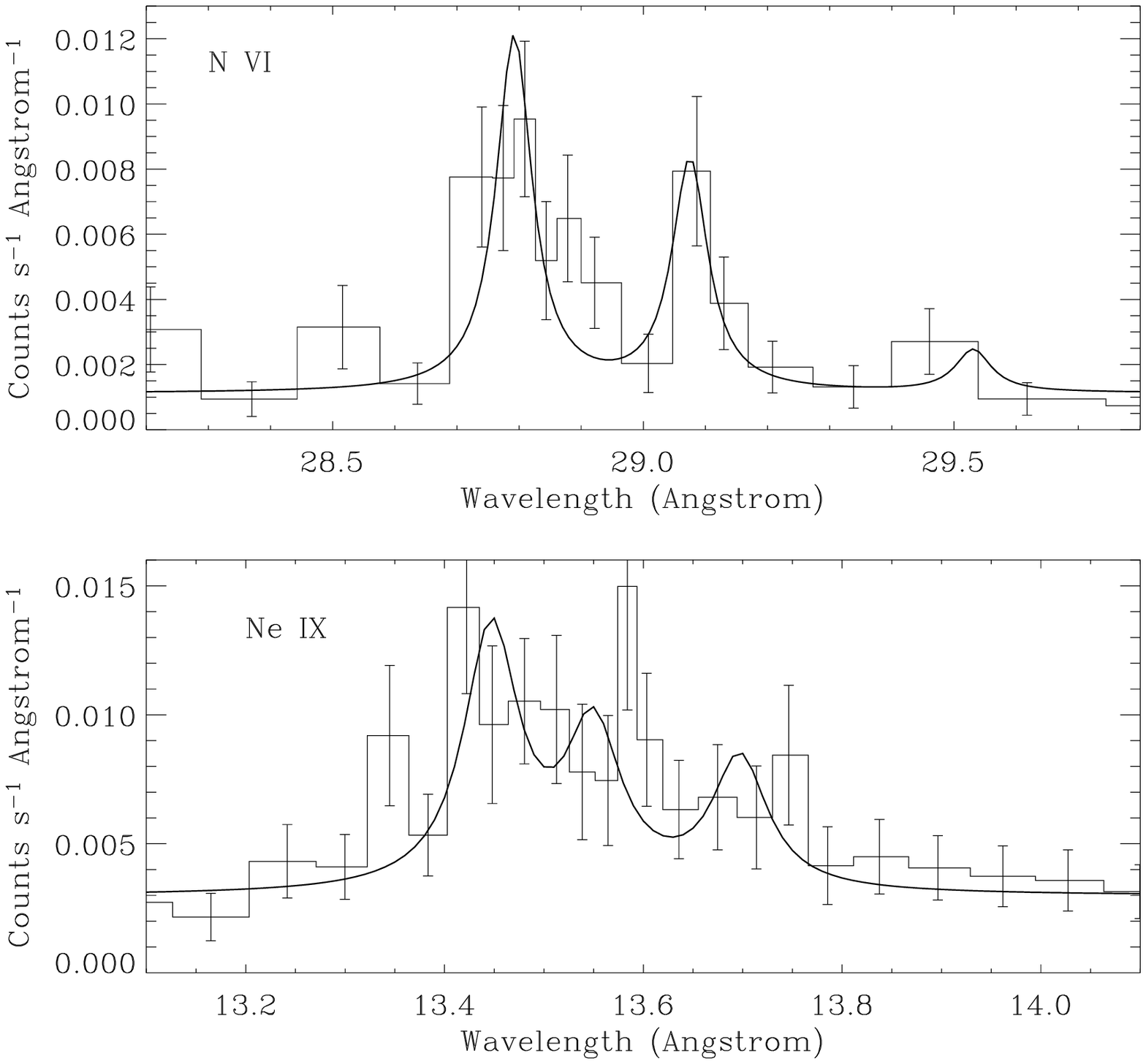}
\end{center}
Figure 3: RGS spectra and best fitting Lorentzian line profiles for
the N VI (top) and Ne IX (bottom) helium-like triplets.
\end{figure}
\clearpage

\begin{figure}
\begin{center}
 \includegraphics[width=6in, height=6in,angle=-90]{f4.ps}
\end{center}
Figure 4: Phase-averaged spectrum of GP Com as observed in the
EPIC/MOS detectors (MOS1:triangles, MOS2:crosses). Shown are the data
and best fitting {\it cevmkl} model versus wavelength using XSPEC. The
fit includes contributions from helium, nitrogen, oxygen, neon,
sulphur and iron.
\end{figure}
\clearpage

\begin{table*}
\begin{center}{Table 1: Emission Line Properties for GP Com}
\begin{tabular}{cccccrcl} \\
\tableline
\tableline
 Species &  &  $\lambda$ ($\AA$) &  &  &  & Flux &  \\
\tableline
  & r & i & f &  &  & ($10^{-4}$ cm$^{-2}$ s$^{-1}$)  &  \\
\tableline
 N VII &  & 24.781 &  &  &  & 1.86 &  \\
 N VII &  & 20.910 &  &  &  & 0.28 &  \\
 N VI  & 28.792 & 29.075 & 29.531 &  & 0.34 & 0.22 & $0.07^1$ \\
 Ne X &  & 12.134 &  &  &  & 0.54 &  \\
 Ne X &  & 10.239 &  &  &  & 0.16 &  \\
 Ne IX  & 13.447 & 13.550 & 13.697 &  & 0.20 & 0.11 & 0.09 \\
\tableline
\end{tabular}
\end{center}
$^1$ Not detected. $1\sigma$ upper limit.
\end{table*}

\clearpage

%%%%%%%%%%%%%%%%%%%%%%%%%%%
%%%%% End of document %%%%%
%%%%%%%%%%%%%%%%%%%%%%%%%%%

\end{document}